\documentclass{iau} 
\usepackage{graphicx}
\usepackage{graphicx}
\usepackage{ulem}
\usepackage{amsmath}
\usepackage{wrapfig}
\usepackage{tikz}
\usepackage{color}

\title[Black hole burning in globular clusters] 
{The Role of ``Black Hole Burning'' in the Evolution of Dense Star Clusters}

\author[Kremer et al.]   
{Kyle Kremer$^{1,\dagger}$,
Claire S. Ye$^1$,
Sourav Chatterjee$^2$,
Carl L. Rodriguez$^3$,
and Frederic A. Rasio$^1$}

\affiliation{$^1$Department of Physics \& Astronomy and Center for Interdisciplinary Exploration \& Research in Astrophysics (CIERA), Northwestern University, Evanston, IL 60208, USA \\[\affilskip]
$^2$ Tata Institute of Fundamental Research, Homi Bhabha Road, Mumbai 400005, India \\[\affilskip]
$^3$ MIT-Kavli Institute for Astrophysics and Space Research, Cambridge, MA 02139, USA \\[\affilskip]
$^\dagger$ email: {\tt kremer@northwestern.edu}
}

\pubyear{2019}
\volume{351}  
\jname{Star Clusters: From the Milky Way to the Early Universe}
\editors{A.~Bragaglia, M.B.~Davies, A.~Sills \& E.~Vesperini, eds.}
\begin{document}

\maketitle

\begin{abstract}
 
  As self-gravitating systems, dense star clusters exhibit a natural diffusion of energy from their innermost to outermost regions, leading to a slow and steady contraction of the core until it ultimately collapses under gravity. However, in spite of the natural tendency toward ``core collapse,'' the globular clusters (GCs) in the Milky Way exhibit a well-observed bimodal distribution in core radii separating the core-collapsed and non-core-collapsed clusters. This suggests an internal energy source is at work, delaying the onset of core collapse in many clusters. Over the past decade, a large amount of work has suggested that stellar black holes (BHs) play a dynamically-significant role in clusters throughout their entire lifetimes. Here we review our latest understanding of BH populations in GCs and demonstrate that, through their dynamical interaction with their host cluster, BHs can naturally explain the distinction between core-collapsed and non-core-collapsed clusters through a process we call ``black hole burning.''

\end{abstract}

\firstsection 
\section{Introduction}
\label{sec:intro}

The study of the evolution of dense star clusters is motivated by the application of these systems to a variety of areas in astrophysics. As high-density environments, star clusters, in particular the old globular clusters (GCs), are expected to facilitate high rates of dynamical encounters, which can lead to the formation of various stellar exotica, including low-mass X-ray binaries, millisecond pulsars, blue stragglers, and cataclysmic variables. Observations of the spatial distribution of GCs in their host galaxies provide constraints on the formation and evolution of galaxies, making GCs valuable tools for extragalactic astronomy. Additionally, over the past several years, GCs have been shown to be efficient factories of the merging binary black hole (BH) systems that may be observed as gravitational-wave sources by LIGO, Virgo, and LISA (e.g., \cite{Banerjee2010,Ziosi2014,Rodriguez2016a,Hurley2016,Chatterjee2017b,Breivik2016,Askar2017,Samsing2018,Kremer2018c}). This, in addition to the discovery of gravitational waves emitted from merging BH binaries by LIGO (\cite{Abbott2016a, Abbott2017}), has sparked renewed interest in understanding the formation and evolution of BHs in GCs.

The old GCs observed in the Milky Way feature a clear bimodality in observed core radius (e.g., \cite{Harris1996,McLaughlin2005}), separating the so-called ``core-collapsed'' clusters from their relatively puffy non-core-collapsed counterparts. Our understanding of the evolution of dense star clusters, and in particular, the dynamical processes that may lead to or prevent core collapse has a long and varied history that has been guided by the complementary efforts of numerical simulations and observations over the past several decades (see, e.g., \cite{Heggie2003} for a thorough review).

Because star clusters are self-gravitating systems with negative heat capacities, dynamical perturbations in a cluster naturally lead to a flow of energy from the strongly self-gravitating core to the relatively sparse halo. The negative heat capacity means that the core becomes even hotter as the result of these perturbations, increasing the flow of energy to the halo in a runaway process that leads to core contraction and ultimately collapse. Thus, the inevitable fate of stellar clusters is to reach a state of so-called ``core collapse." This process was first demonstrated in the 1960s by \cite{Antonov1962} and \cite{Lynden-Bell1968} and, around the same time, \cite{Henon1961} constructed the first
cluster models exhibiting core collapse.

However, many of the old GCs observed in the Milky Way are very clearly not core collapsed (see, e.g., \cite{Harris1996} and also see \cite{Elson1991} and \cite{Mackey2008} for similar discussion of the Magellanic Cloud clusters). Many clusters have very large well-resolved cores and luminosity profiles that are well-fit by King models (\cite{King1962}) and show no signs of the central cusp that we would expect from the onset of this instability. This implies that there exists some internal energy source which balances the outward flow of energy and supports the cores against this tendency toward collapse.

For the past several decades, binary stars have been studied as a plausible explanation for halting cluster core collapse.\footnote{Other mechanisms have also been proposed including a massive central black hole and mass loss from the stellar evolution of merger products. We do not discuss these further here but see, e.g., \cite{Goodman1989} for a review.} For some time, binaries in GCs were thought to exclusively form dynamically through three-body interactions (e.g., \cite{Heggie2003}); however, since the early 1990s, motivated by HST observations, theoretical analyses have focused on studying properties of clusters with primordial binary populations as they pass through the so-called ``binary-burning'' phase, where the cluster core is supported against collapse by super-elastic dynamical scattering interactions of binary stars (e.g.,\cite{Vesperini1994,Fregeau2007,Chatterjee2013a}). 

\begin{figure}
\begin{center}
\includegraphics[width=0.8\linewidth]{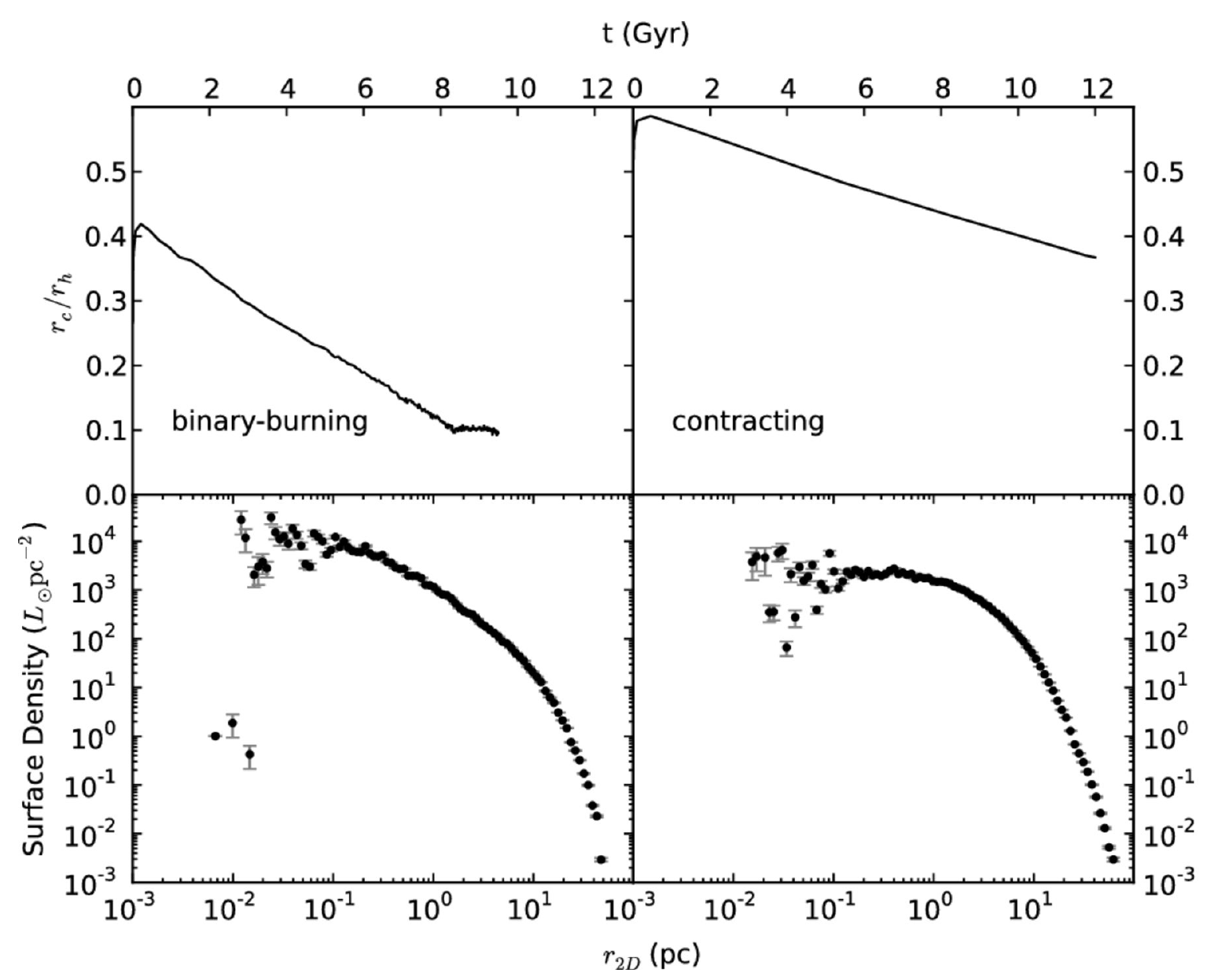}
\caption{Examples of a binary-burning cluster (left) and a core-contracting cluster (right) chosen from a set of \texttt{CMC} cluster models. The top panels show the evolution of $r_c/r_h$ for each cluster. The bottom panels show the surface luminosity density profiles. The binary-burning model shows a clear power-law slope until the data are too noisy. In contrast, the core-contracting model shows a clear King density profile. This is a reproduction of Figure 11 from ``Understanding the dynamical state of globular clusters: core-collapsed versus non-core-collapsed,'' Chatterjee et al. 2013, MNRAS, 429, 2881, DOI: 10.1093/mnras/sts464. Reproduced with permission.}
\end{center}
\label{fig:binburning}
\end{figure}

However, more recent work, in particular \cite{Chatterjee2013a}, has demonstrated that the connection between binary burning and core collapse (or lackthereof) is more subtle. In Figure \ref{fig:binburning}, we show two characteristic models, one that reaches binary-burning equilibrium and one that does not. For the model on the left, the evolution of the core radius flattens out at a time of about 8 Gyr, indicating that binary burning has begun. But examination of the luminosity profile of this model reveals that once binary burning has started, the cluster is clearly representative of a core-collapsed cluster: there is a prominent cusp at small $r$ and a very tiny core. On the other hand, on the right we see a cluster that is still contracting at present day; it has yet to reach binary-burning equilibrium and it very clearly has a luminosity profile with a large well-resolved core that exhibits a clear King density profile. In other words, binary-burning clusters in thermal equilibrium have such tiny cores and centrally condensed density profiles that they would almost certainly be observationally identified as ``core-collapsed'' clusters. The question remains as to what is the energy source acting in clusters during the core-contraction phase. If not binary burning, what is the mechanism that delays the onset of core collapse in the majority of clusters observed in the Milky Way?

Arguably, the most important recent shift in our understanding of how GCs evolve came from the observational and theoretical confirmation that GCs contain dynamically-important populations of stellar-mass BHs up to the present time. Being the most massive objects in a GC, the BH population collapses quickly and generates energy through dynamical binary formation, BH-binary burning, and dynamical ejections (see Section \ref{sec:BHs} for details and references). Through strong dynamical encounters in the inner, BH-dominated region, BHs are frequently ejected to higher orbits in the cluster potential, leading to interactions with luminous stars in the outer parts of the cluster. Through these interactions, the BHs deposit energy into the GC's stellar bulk and influence the large-scale structural properties of their host cluster (e.g., \cite{Merritt2004,Mackey2007,Mackey2008,BreenHeggie2013,Peuten2016,Wang2016,ArcaSedda2018,Kremer2018d,Kremer2019a}). Specifically, while present in GCs, BHs produce energy through a process we label ``BH burning'' at a rate sufficient to delay the collapse of the cluster's core. During BH burning, clusters are in the core-contraction phase, similar to the cluster shown in the right-hand panel of Figure \ref{fig:binburning}. As shown in Figure \ref{fig:binburning}, core-contracting (as opposed to core-\textit{collapsed}) clusters are characterized by surface brightness profiles representative of the majority of clusters in the Milky Way (i.e., can be well-modeled by King profiles). \textit{Only after the stellar-mass BH population is significantly depleted}, can the surface brightness profile of a GC reach a traditional core-collapse architecture (e.g., \cite{Kremer2018d}). At this stage, in the absence of a large number of BHs, the luminous binaries become the dominant source of energy at the GC's center and the well-studied, traditional binary-burning phase begins.

\section{Black Holes in Globular Clusters}
\label{sec:BHs}

Thousands of BHs are likely to form in GCs as the result of the evolution of massive stars. The number of these BHs that are \textit{retained} in GCs today is less certain. BHs are expected to be ejected from their host GCs through one of two primary mechanisms: ejection due to sufficiently large natal kicks or ejection via recoil as a result of strong dynamical encounters with other remaining BHs.

BH natal kicks, which are caused by asymmetric mass loss of supernova ejecta,
are poorly constrained (e.g., \cite{Belczynski2002,Fryer2012,Mandel2016,Repetto2017}). If BH natal kicks are comparable in magnitude to the high speeds expected for the natal kicks of core-collapse neutron stars (NSs; e.g., \cite{Hobbs2005}), the vast majority of BHs are likely to be ejected from GCs immediately upon formation because of the low escape speeds of typical GC cores. However, in the case of weaker BH natal kicks, a potentially large fraction of BHs may be retained post-supernova.

The long-term retention of BHs that are not ejected promptly from natal kicks has long been a subject of debate. Until relatively recently, it was argued that BHs retained after formation would quickly mass-segregate and form a dense sub-cluster dynamically decoupled from the rest of the GC (e.g., \cite{Spitzer1967,Kulkarni1993,Sigurdsson1993}). The BH members of this compact sub-cluster would then undergo strong dynamical encounters, ultimately ejecting all but a few BHs from the cluster on sub-Gyr timescales. However, more recently, several theoretical and computational analyses have demonstrated that this argument of rapid BH evaporation is not correct, and in fact, many BHs may be retained at present (e.g., \cite{Merritt2004,Hurley2007,Mackey2007,Mackey2008,Morscher2015,Kremer2019a}).

The topic of retained BHs in GCs has been further motivated observationally. In the past decade, several stellar-mass BH candidates have been identified in both Galactic (\cite{Strader2012,Chomiuk2013,Miller-Jones2014,Shishkovsky2018}) and extragalactic (\cite{Maccarone2007,Irwin2010}) GCs. Most recently, the first stellar-mass BH to be identified through radial velocity measurements was found in the Milky Way GC NGC 3201 (\cite{Giesers2018}). The observations of these stellar-mass BH candidates suggest that at least some GCs do indeed retain populations of BHs at present and given that these host GCs do not show any particular trends in their observable properties, it appears that BH retention to present day may be common to most GCs. 

In general, the natal kick strengths determine the fraction of BHs retained immediately post formation. Subsequently, the cluster ejects BHs via dynamical processing including mass segregation and strong scattering over several relaxation times. Properties of the host cluster at birth determine the rate at which these dynamical processes act. Observations of young massive clusters (e.g., \cite{Bastian2005,Fall2005,Gieles2006,Scheepmaker2007,Scheepmaker2009,PortegiesZwart2010}), the expected progenitors of GCs, can provide insight into the various initial cluster properties that may determine the eventual outcome of the cluster, in particular, the rate at which a cluster's BHs are dynamically ejected and thus, whether or not the GC has undergone core collapse by the present day. In the next section, we describe how a cluster's initial size is the key parameter relevant to young massive clusters that determines the long-term fate of the cluster and its BH population.


\newpage
\section{How Black Holes Shape Globular Clusters}

From the perspective of modelling GCs, the initial cluster size can be specified in terms of the cluster virial radius, $r_v$, a theoretical quantity defined as

\begin{equation}
r_v = \frac{GM^2}{2\lvert U \rvert}
\end{equation}
where $M$ is the total cluster mass and $U$ is the total cluster potential energy. The initial relaxation timescale is directly related to the initial cluster size. Thus, the initial $r_v$ sets the dynamical clock of each cluster and controls how dynamically old a particular cluster is at a fixed physical time window, which, in turn, determines how close or far the cluster is from undergoing core collapse. The half-mass relaxation time is given by

\begin{equation}
\label{eq:relaxation_time}
t_{\rm{rh}} = 0.138 \frac{M^{1/2}R_{\rm{h}}^{3/2}}{\langle m \rangle G^{1/2}\ln \Lambda}
\end{equation}
(Equation 2-63 \cite{Spitzer1987}), where $R_{\rm{h}}$ is the half-mass radius, $\langle m \rangle$ is the mean stellar mass, and $\ln \Lambda$ is the Coloumb logarithm where $\Lambda \simeq 0.4 N$, where $N$ is the total number of particles.

\cite{Kremer2019a} recently demonstrated that the evolution of stellar-mass BH populations in GCs is intimately related to the contraction or expansion of the GC's core radius. In particular, since the initial $r_v$ of a cluster determines the initial relaxation timescale, the initial $r_v$ also controls how dynamically processed the BHs are at any given late physical time. As a result, by exploring a range in initial cluster sizes motivated by observations of young massive clusters, the full spectrum of GC types at the present-day, ranging from core-collapsed clusters to puffy clusters with large observed core radii, is naturally produced.

\begin{figure}
\begin{center}
\includegraphics[width=0.6\linewidth]{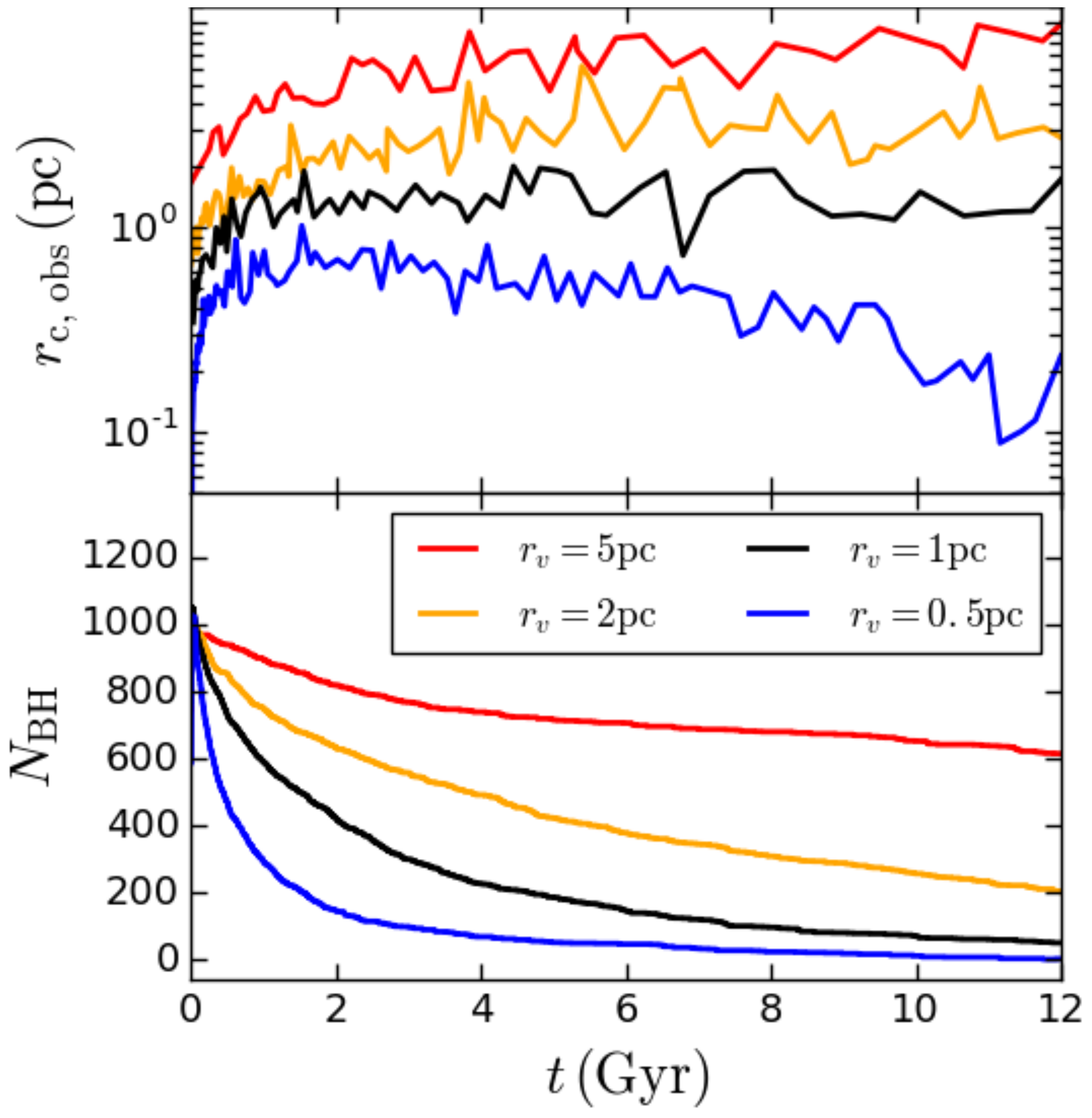}
\label{fig:rc}
\caption{Observed core radius (top panel) and total number of BHs (bottom panel) versus time for models with four different initial virial radii: $r_v=5$pc (red curve), $r_v=2$ pc (orange curve), $r_v = 1$ pc (black), and $r_v = 0.5$ pc (blue). Note that models with larger initial virial radii (longer relaxation timescales) retain relatively more BHs and exhibit larger core radii at present compared to models with smaller initial virial radii (shorter relaxation timescales.) This is a direct consequence of the transport of energy from the BH core to the outer parts of the cluster or, as we refer to it here, ``BH burning." Adapted from Figure 2 of ``How Initial Size Governs Core Collapse in Globular Clusters,'' Kremer et al. 2019, ApJ, 871, 38, DOI: 10.3847/1538-4357/aaf646.}
\end{center}
\end{figure}

To illustrate these processes, we show data taken from a set of 11 fully-evolved cluster models run with the \texttt{Cluster Monte Carlo} code (\texttt{CMC}) in \cite{Kremer2019a}. \texttt{CMC} is a fully-parallelized code that uses H\'{e}non-style Monte Carlo methods to model the long-term evolution of GCs (for a review, see \cite{Henon1971a,Joshi2000,Joshi2001, Fregeau2003, Umbreit2012, Pattabiraman2013, Chatterjee2010, Chatterjee2013a,Rodriguez2018}). \texttt{CMC} uses the stellar evolution packages \texttt{SSE} (\cite{Hurley2000}) and \texttt{BSE} (\cite{Hurley2002}) to model the evolution of single stars and binaries and uses the \texttt{Fewbody} package (\cite{Fregeau2004,Fregeau2007}) to model the evolution of three- and four-body encounters. \texttt{CMC} has been developed over the past decade-plus and has been shown to agree well with the results of $N$-body simulations of GCs. For a review of the most up-to-date modifications to \texttt{CMC}, including the incorporation of post-Newtonian terms into all few-body encounters, see \cite{Rodriguez2018}. 

For this set of 11 models, a number of initial parameters are fixed including: total particle number, $N=8 \times 10^5$; King concentration parameter, $w_o = 5$; binary fraction, $f_b=5\%$; metallicity, $\rm{Z}=0.001$; and Galactocentric distance, $d=8$ kpc. The initial mass function for all stars and the initial period distribution for all binaries are chosen as in \cite{Kremer2018d}. We adopt the prescription for stellar remnant formation described in \cite{Fryer2001} and \cite{Belczynski2002}. Natal kicks for core-collapse NSs are drawn from a Maxwellian with dispersion width $\sigma_{\rm{NS}} = 265 \rm{km\,s}^{-1}$ (\cite{Hobbs2005}). We assume BHs are formed with fallback and calculate the BH natal kicks by sampling from the same kick distribution as the NSs, but with the BH kicks reduced in magnitude according to the fractional mass of fallback material (see \cite{Morscher2015} for more details).

The only parameter varied within this set of models is the initial virial radius, $r_v$, between values of $0.5-5\,$pc. As described in Section \ref{sec:BHs}, the initial virial radius directly determines the cluster's initial relaxation time, and therefore, determines the rate at which BHs are ejected from the cluster. This is illustrated in Figure 2, where we show the evolution of the observed core radius (top panel) and total number of retained BHs (bottom panel) as a function of time for four models that bracket the full range in initial virial radii considered in the study. As the figure clearly shows, models with smaller initial virial radii (shorter initial half-mass relaxation times; see Equation 3.2) eject their BHs faster compared to models with larger initial virial radii (longer initial relaxation times). By adjusting the initial virial radii by only one order-of-magnitude (and keeping all other initial cluster parameters fixed), the number of BHs retained in the cluster at $t=12\,$Gyr varies from 0 to over 600. In response to the energy produced by the cluster's BH population at late times, models with few BHs have relatively small cores compared to models which still retain large BH populations ($N_{\rm{BH}} \gtrsim 100$).

\begin{figure}
\begin{center}
\includegraphics[width=\linewidth]{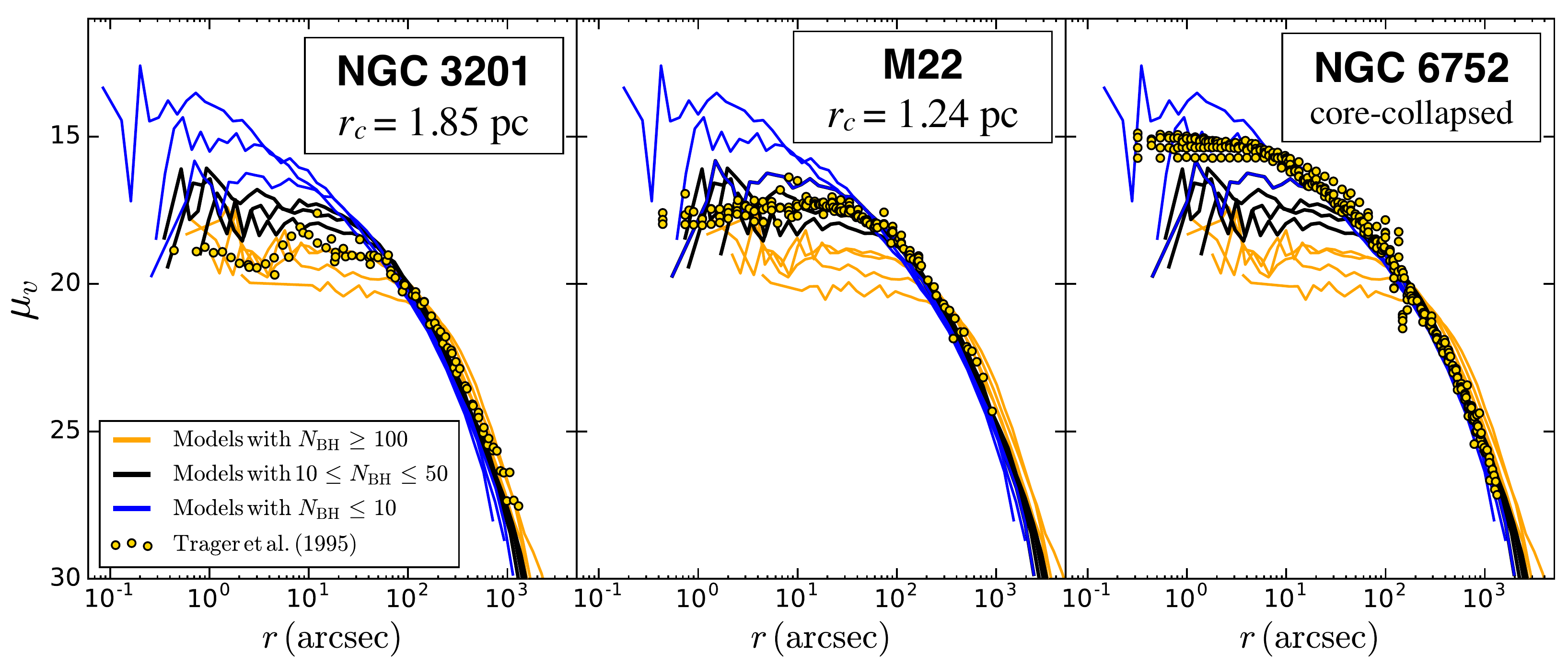}
\caption{Surface brightness profiles (SBPs) at $t=12$ Gyr for a set of 11 models with initial virial radii ranging from $0.5-5\,$pc. Orange curves denote BH-rich models ($N_{\rm{BH}} \geq 100$ at $t=12$ Gyr), blue curves denote BH-poor models ($N_{\rm{BH}} \leq 10$), and black curves denote models with intermediate number of BHs. The left-hand panel shows model SBPs compared to the observed SBP for NGC 3201 (gold circles). The middle panel shows the models compared to M22, and the right compares to the core-collapsed cluster NGC 6752. All observed SBPs are taken from \cite{Trager1995}. Note the general trend that clusters with large core radii are better matched by models with large numbers of BHs at present, while core-collapsed clusters are best matched by models in which the BH population has been depleted. Adapted from Figure 4 of ``How Initial Size Governs Core Collapse in Globular Clusters,'' Kremer et al. 2019, ApJ, 871, 38, DOI: 10.3847/1538-4357/aaf646.}
\end{center}
\label{fig:SBP}
\end{figure}

In Figure \ref{fig:SBP}, we illustrate the effect of BH burning in a slightly different way. Here, we show surface brightness profiles (SBPs) for each of these models at 12 Gyr compared to the observed SBP (from \cite{Trager1995}) for several Milky Way GCs: NGC 3201 (left), M22 (middle), and NGC 6752 (right). Stellar-mass BH candidates have been identified in both NGC 3201 (\cite{Giesers2018}) and M22 (two candidates; \cite{Strader2012}), while NGC 6752 is a well-observed core-collapsed cluster. Importantly, all three clusters have roughly comparable total mass, which allows us to attribute differences solely to the cluster's dynamical evolution. Together, these three cluster roughly span the full range in cluster types (from the ``core-collapsed'' to ``puffy'' extremes). 

Here the models are divided by color into three categories: BH-rich models (defined as $N_{\rm{BH}} \geq 100$; orange curves), BH-poor models (defined as $N_{\rm{BH}} \leq 10$; blue curves), and models with intermediate numbers of BHs (defined as $15 < N_{\rm{BH}} < 50$; black curves). As Figure \ref{fig:SBP} clearly shows, BH-rich models produce clusters most similar to NGC 3201 at late times, while BH-poor models feature SBPs with prominent cusps at low $r$, representative of core-collapsed MW GCs, such as NGC 6752. Between the puffy and core-collapse extremes, we have models with intermediate number of BHs, that most accurately match the SBPs of M22.

Figure \ref{fig:SBP} shows a result similar to Figure 2: for clusters with similar total mass, a clear correlation exists between the size of a cluster's core and the size of its BH population. Furthermore, the size of the BH population is directly linked to the initial size of the cluster: the initial size (here, $r_v$), sets the cluster's half-mass relaxation time which determines how quickly the BHs are depleted. For cluster that still retains a dynamically-significant population of BHs at present day, the BHs dominate the energy balance of the cluster through BH burning, preventing the collapse of the cluster's luminous core. On the other hand, for clusters born with relatively higher densities (smaller initial $r_v$), the BH population is relatively depleted at late times and the energy production from the BH core is insufficient to prevent collapse of the luminous core, and the cluster reaches a core-collapse state.

\section{Discussion and Conclusions}

In the past decade, major strides have been made in our understanding of the formation and evolution of stellar-mass BH populations in dense star clusters. As recently as twenty years ago, it was widely believed that realistic GCs would retain zero to only a few BHs at present. However, recent observational and computational advances have provided myriad evidence that not only are BHs present in some GCs at present, but they play an essential role in the dynamical evolution of their host cluster throughout the cluster's full lifetime. To this end, recent work has demonstrated that stellar-mass BHs may provide the answers to several key questions in the field of star cluster dynamics. Through a process we refer to as ``BH burning'' (in analogy with the classic ``binary-burning" process), a number of analyses have shown that a centrally-concentrated population of BHs can satisfy the energy demands of the whole cluster, preventing the collapse of the cluster's core. Thus, BH burning yields a natural explanation for the observed bimodal distribution in core radii observed in the Milky Way GCs: While BH burning is efficient, the core remains expanded and the cluster appears to be non-collapsed. Only when BH burning becomes inefficient due to sufficient BH ejections, does the core collapse. At this point, classical luminous binary burning starts and the cluster attains properties similar to a core-collapsed cluster.

We have reviewed a number of recent papers demonstrating this process, in particular, \cite{Kremer2019a}. This analysis shows that the key parameter for determining the evolution of a cluster's BH population, and thus, the evolution of the cluster as a whole, is the cluster's initial size. \cite{Kremer2019a} demonstrated that by exploring a small range in initial cluster size (parameterized in terms of the initial cluster virial radius) motivated by observations of young massive clusters, the full spectrum of present-day cluster types, ranging from core-collapsed to non-core-collapsed, is naturally produced. Furthermore, this analysis shows that several Milky Way GCs (in particular clusters such as NGC 3201, M10, and M22 that host observed BH candidates) have structural features consistent with large BH populations at present while others (such as the core-collapsed cluster NGC 6752) are more consistent with models that retain zero to few BHs at present. This suggests, for clusters with similar total mass, a clear correlation between the size of a cluster's core and its BH population.


Of course, there remains ample room for future work in this field. For example, a better theoretical understanding of the dynamical formation processes for various stellar exotica in clusters such X-ray binaries, millisecond pulsars, blue stragglers, and cataclysmic variables will allow more detailed comparisons to be made between cluster models and observations. In particular, through the same mechanisms that BHs influence the properties of their host cluster, BH populations also likely influence the formation of these various objects indirectly. For example, recent work by \cite{Ye2018} has already suggested that the formation of millisecond pulsars in GCs is intimately tied to a cluster's BHs. Thus, objects such as millisecond pulsars and other stellar exotica may be used as additional observational tracers of BH populations in clusters. 

Furthermore, the process through which GCs are formed is highly complex. The first few to 10s of Myrs of cluster evolution likely feature various complex processes such as hierarchical mergers and residual gas expulsion. Indeed, such processes are hinted at from observations of several young massive clusters (e.g., \cite{Kuhn2014,Gennaro2017}). In particular, residual gas expulsion could lead to significant cluster expansion at early times, attenuating the long-term dynamical processing of BHs and thus altering the structural features of the clusters at late times. It remains unclear how inclusion of the various processes relevant to cluster formation and the earliest stages of cluster evolution influence both the formation of BH populations in clusters and the symbiotic relation between BHs and their host cluster that we have discussed here. More detailed studies examining these various processes are needed.

\end{document}